\documentclass[superscriptaddress,reprint,aps,prl]{revtex4-1}
\usepackage{amssymb}
\usepackage{amsmath}
\usepackage{graphicx}
\usepackage{dcolumn}
\usepackage{bm}
\usepackage{float}
\usepackage{xcolor}
\usepackage[normalem]{ulem}

\begin{document}

\title{Crystal fields and magnetic structure of the Ising antiferromagnet Er$_3$Ga$_5$O$_{12}$}

\author{Y. Cai}
\author{M.N. Wilson}
\author{J. Beare}
\author{C. Lygouras}
\author{G. Thomas}
\address{Department of Physics and Astronomy, McMaster University, Hamilton, ON L8S 4M1, Canada}
\author{D.R. Yahne}
\address{Department of Physics, Colorado State University, Fort Collins, CO 80523, USA}
\author{K. Ross}
\address{Department of Physics, Colorado State University, Fort Collins, CO 80523, USA}
\address{Canadian Institute for Advanced Research, Toronto, ON M5G 1M1, Canada}
\author{K.M. Taddei}
\address{Neutron Scattering Division, Oak Ridge National Laboratory, Oak Ridge, TN 37831, USA}
\author{G. Sala}
\address{Neutron Scattering Division, Oak Ridge National Laboratory, Oak Ridge, TN 37831, USA}
\author{H.A. Dabkowska}
\address{Brockhouse Institute for Materials Research, McMaster University, Hamilton, ON L8S 4M1, Canada}
\author{A.A. Aczel}
\email{aczelaa@ornl.gov}
\address{Neutron Scattering Division, Oak Ridge National Laboratory, Oak Ridge, TN 37831, USA}
\address{Department of Physics and Astronomy, University of Tennessee, Knoxville, TN 37996, USA}
\author{G.M. Luke}
\email{luke@mcmaster.ca}
\address{Department of Physics and Astronomy, McMaster University, Hamilton, ON L8S 4M1, Canada}
\address{Canadian Institute for Advanced Research, Toronto, ON M5G 1M1, Canada}
\address{TRIUMF, Vancouver, BC V6T 2A3, Canada}

\begin{abstract}
Rare earth garnets are an exciting playground for studying the exotic magnetic properties of the frustrated hyperkagome lattice. Here we present a comprehensive study of the single ion and collective magnetic properties of the garnet Er$_3$Ga$_5$O$_{12}$. Using inelastic neutron scattering, we find a crystal field ground state doublet for Er$^{3+}$ with strong Ising anisotropy along local [100] axes. Magnetic susceptibility and heat capacity measurements provide evidence for long-range magnetic ordering with $T_N$~$=$~0.8~K, and no evidence for residual entropy is found when cooling through the ordering transition. Neutron powder diffraction reveals that the ground state spin configuration corresponds to the six-sublattice, Ising antiferromagnetic state ($\Gamma_3$) common to many of the rare earth garnets. However, we also found that $\mu$SR appears to be insensitive to the ordering transition in this material, in which a low-temperature relaxation plateau was observed with no evidence of spontaneous muon precession. The combined muon and neutron results may be indicative of a dynamical ground state with a relatively long correlation time. Despite this potential complication, our work indicates that Er$_3$Ga$_5$O$_{12}$ is an excellent model system for studying the complex metamagnetism expected for a multi-axis antiferromagnet.
\end{abstract}

\maketitle

\section{Introduction}
Frustrated magnets are materials in which the competing pairwise interactions between magnetic moments cannot be satisfied simultaneously. The pyrochlore lattice, which consists of corner-sharing tetrahedra, represents one of the canonical three-dimensional (3D) frustrated geometries\cite{06_greedan, 10_gardner, 15_wiebe}. Among this large family of materials, the rare earth pyrochlores of the form $R_2 B_2$O$_7$ ($R$~$=$~rare earth, $B$~$=$~non-magnetic cation) have received the most attention due to their rich variety of magnetic ground states and excitations. This large variation between states, arising from the interplay of exchange couplings, dipolar interactions, and single ion anisotropy, leads to spin glasses\cite{91_reimers, 14_silverstein}, spin liquids\cite{17_wen, 19_gao, 19_gaudet}, spin ices\cite{01_snyder}, order-by-disorder\cite{03_champion, 14_ross}, magnetic moment fragmentation\cite{16_benton, 16_petit}, and conventional long-range magnetic ordering.  

\begin{figure}
\includegraphics[width=\columnwidth]{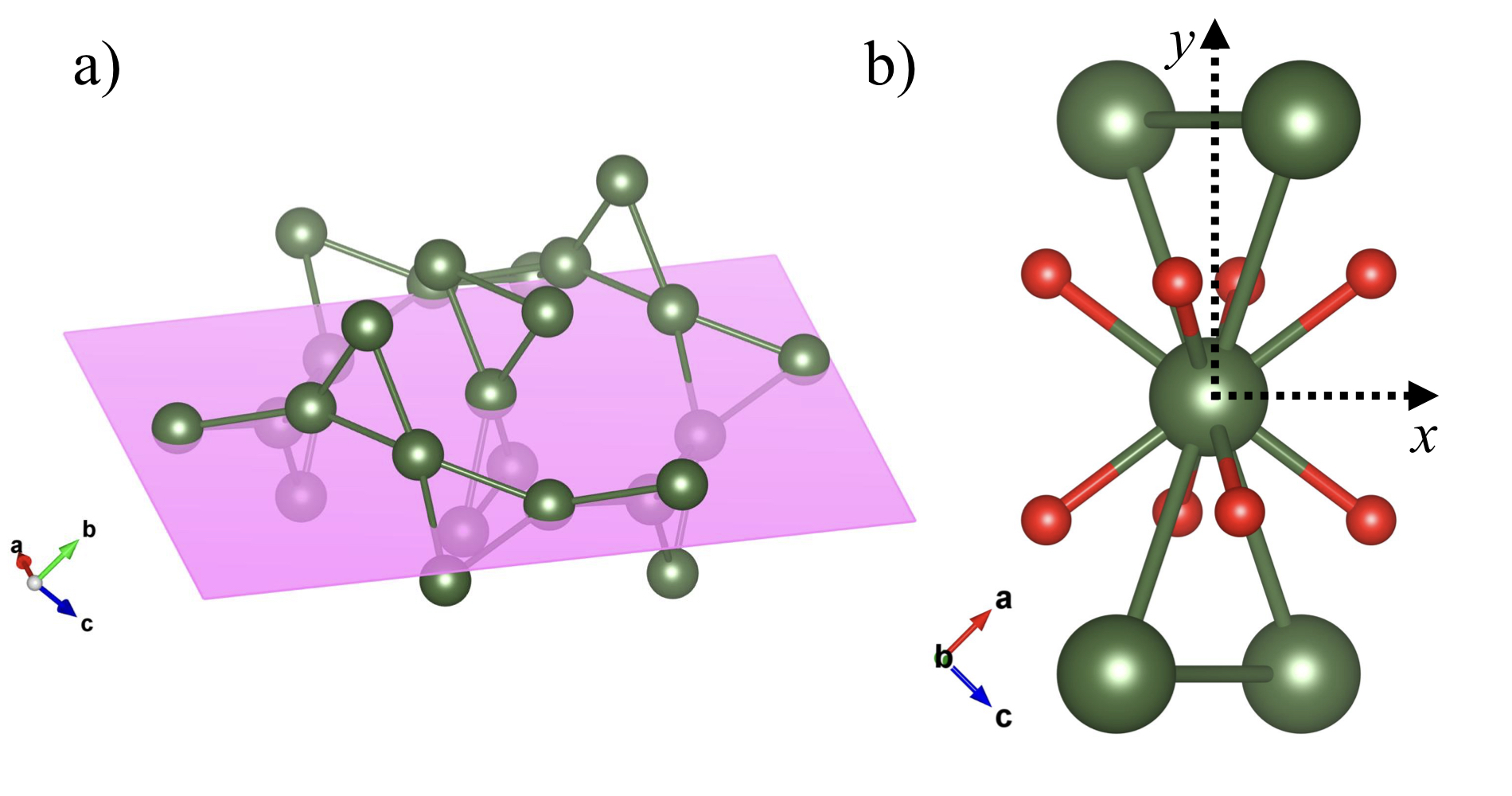}
\caption{(a) Corner-sharing triangular network of Er$^{3+}$ ions in Er$_3$Ga$_5$O$_{12}$; the 0{$\overline1$}1 crystallographic plane is shown in pink. (b) The local environment of an Er$^{3+}$ ion in Er$_3$Ga$_5$O$_{12}$, this schematic is appropriate for the ion at site (0.25, 0.125, 0). Each Er$^{3+}$ ion has $D_{2}$ orthorhombic point symmetry and is surrounded by eight O$^{2-}$ ions.}
\label{Fig: Er_network}
\end{figure}

Another common 3D frustrated architecture, based on corner-sharing triangles, is the hyperkagome lattice. This geometry is realized by the rare earth sublattice in the garnets $R_3$(Ga,Al)$_5$O$_{12}$, and is illustrated in Fig.~\ref{Fig: Er_network}(a). The most commonly studied material in this family is the Heisenberg garnet Gd$_3$Ga$_5$O$_{12}$, as it hosts a spin liquid state above a freezing temperature of $T_g$~$=$~0.14~K\cite{79_kinney, 98_petrenko, 00_dunsiger, 02_marshall, 06_yavorskii, 10_deen, 13_quilliam, 15_dambrumenil, 15_deen} that has recently been argued to arise from hidden multipolar order\cite{15_paddison}. A Kagome spin ice state with extensive degeneracy has also been proposed for Ising garnets when the moments lie along the local [110] axes\cite{04_yoshioka}, but this exotic state has yet to be uncovered in the laboratory. Finally, complex metamagnetic behavior has been predicted and observed for garnets with strong Ising anisotropy along local [100] axes\cite{66_keen, 66_keen_2, 81_felsteiner, 82_steiner}.

Establishing the hiearchy of interactions in $R_3$(Ga,Al)$_5$O$_{12}$ is an important step towards gaining a detailed understanding of the magnetic properties of these materials. As in the case of rare earth pyrochlores, exchange couplings, dipolar interactions, and single ion anisotropy arising from crystal fields are all expected to be important. While the single ion anisotropy is typically the easiest contribution to quantify, surprisingly little is known with certainty for this class of materials. Most previous crystal field studies were performed using optical spectroscopy since these materials were first investigated back in the 1960s-1970s\cite{63_wong, 64_dreyfus, 66_hooge, 66_koningstein, 67_buchanan, 70_johnson}, although it is not always straightforward to differentiate between phonons and low-lying crystal field levels with this technique. It is now understood that inelastic neutron scattering is the premier method for measuring crystal field excitations, as they can be unambiguously identified due to a momentum transfer ($Q$) dependence that is different from phonons. Despite this significant advantage, only the crystal field parameters of Ho$_3$Ga$_5$O$_{12}$\cite{91_reid} have been determined with inelastic neutron scattering to-date.

The collective magnetic properties of many rare earth aluminum and gallium garnets were also investigated several decades ago, typically with a combination of magnetic susceptibility, heat capacity, and neutron diffraction measurements. Many of these materials achieve six-sublattice antiferromagnetic long-range order corresponding to the $\Gamma_3$ irreducible representation in Kovalev's notation\cite{kovalev} at temperatures $T$~$<$~3~K\cite{65_hastings, 69_hammann, 77_hammann}. While this ordered state is consistent with expectations for dipolar interactions between local [100] Ising moments\cite{65_capel}, the collective magnetic properties of several systems remain poorly understood. For instance, there is a sharp $\lambda$ anomaly in the specific heat of Yb$_3$Ga$_5$O$_{12}$ at $T_N$~$=$~54~mK\cite{03_dalmas} initially thought to be indicative of long-range magnetic order, but muon spin relaxation\cite{03_dalmas} and Mossbauer spectroscopy\cite{03_hodges} measurements show no evidence for the expected order. Neutron diffraction has also revealed that the magnetic ground state of Nd$_3$Ga$_5$O$_{12}$ corresponds to the $\Gamma_4$ irreducible representation\cite{68_hammann}, which is still consistent with local [100] Ising moments but not predicted by Capel's original theory\cite{65_capel}. Finally, revisiting the collective magnetic properties of these materials with modern neutron scattering instrumentation has led to some surprises. In particular, Ho$_3$Ga$_5$O$_{12}$ was initially reported to order in the $\Gamma_3$ magnetic structure\cite{77_hammann}, but recent neutron scattering measurements found the coexistence of diffuse magnetic scattering with the $\Gamma_3$ magnetic Bragg peaks in zero field, with a reasonably small field of 20~kG being enough to suppress the diffuse scattering\cite{08_zhou}. This finding provides strong motivation to reexamine the collective magnetic properties of other Ising garnets, as they may also be more complex than first proposed in the earlier studies.

In this work, we address the magnetism in Er$_3$Ga$_5$O$_{12}$ using multiple techniques. Using inelastic neutron scattering, we find a doublet crystal field ground state for Er$^{3+}$ with strong Ising anisotropy along local [100] axes. We also investigate the collective magnetic properties of this system with magnetic susceptibility, heat capacity, and neutron diffraction measurements, which reveal $\Gamma_3$ antiferromagnetic ordering with $T_N$~$=$~0.8~K. No diffuse scattering is observed below $T_N$ and no residual entropy is found when cooling through the ordering transition, which are both consistent with a conventional ordered state.

\section{Crystal growth and experimental details}

Single crystals of Er$_3$Ga$_5$O$_{12}$ were grown by the traveling-solvent floating-zone (TSFZ) technique at McMaster University. Stoichiometric mixtures of high-purity raw materials Er$_2$O$_3$ (99.99\%) and Ga$_2$O$_3$ (99.999\%) were ground, pressed hydrostatically into rods and heated in air at 1200$^\circ$C for 8 hours. No sample mass loss was observed before or after heating. Crystal growth was performed at a growth rate of 5~mm/hr and 1 atm overpressure under Argon gas, resulting in a large, transparent, pink crystal with a length of 3~cm. The crystals were determined to be single phase Er$_3$Ga$_5$O$_{12}$ by performing powder x-ray diffraction measurements on a crushed portion of the crystals. We collected magnetic susceptibility measurements from 0.48~K to 300~K on a small cut single crystal with the magnetic field applied along the [110] direction using a Quantum Design Magnetic Property Measurement System XL-3 equipped with an iQuantum He$^3$ Insert for measurements below 2~K. We also measured the specific heat of a single crystal between 0.1~K and 4~K using a Quantum Design Physical Property Measurement System with a dilution fridge insert. 

The inelastic neutron scattering (INS) experiment to measure the crystal field levels was performed using the SEQUOIA spectrometer\cite{10_granroth} at the Spallation Neutron Source of Oak Ridge National Laboratory (ORNL) using 9.4~g of crushed single crystals loaded into an Al cylindrical can. All data were collected with incident energies $E_i$~$=$~30~meV and 120~meV, with corresponding fine Fermi chopper frequencies of 240~Hz and 600~Hz, resulting in instrumental energy resolutions of 0.65~meV and 2.1~meV (full-width half-maximum) respectively at the elastic line. A closed cycle refrigerator was used to achieve a base temperature of 4~K. The neutron powder diffraction (NPD) experiment to investigate the magnetic structure was carried out on the HB-2A powder diffractometer\cite{18_calder} at the High Flux Isotope Reactor of ORNL using the same sample measured in the SEQUOIA experiment. A cryostat with a He$^3$ insert was used to achieve a base temperature of 0.3~K. A collimation of open-21$'$-12$'$ was used in this experiment, and the FULLPROF software suite\cite{93_rodriguez} was used to perform all structural and magnetic refinements reported in this work. 

Muon spin relaxation measurements were performed at TRIUMF, Canada on the the M20 and M15 beamlines with He$^4$ gas flow cryostat (base $T$~$=$~1.5 K) and dilution fridge (base $T$~$=$~30~mK) setups. A single crystal of Er$_3$Ga$_5$O$_{12}$ was mounted on the M20 beamline with a low background apparatus in the He$^4$ cryostat using aluminum backed mylar tape. Another single crystal sample was sliced into $\sim$1 mm thick discs and mounted onto an Ag plate and covered in thin Ag foil for the measurements on the M15 beamline. All measurements were performed with zero applied field (ZF) and all the $\mu$SR data were fit by the open source $\mu$SRfit software package\cite{suter}.

\section{Results and discussion}

\subsection{Inelastic neutron scattering}

In Er$_3$Ga$_5$O$_{12}$, the Er$^{3+}$ ions occupy two interpenetrating, corner-sharing triangular sublattices as shown in Fig.~\ref{Fig: Er_network}(a). Each Er$^{3+}$ ion is surrounded by eight nearest-neighbor oxygen ions, leading to dodecahedral local geometry and $D_{2}$ orthorhombic point symmetry as shown in Fig.~\ref{Fig: Er_network}(b) for Er site (0.25, 0.125, 0). According to Hund's rules, the ground state multiplet of an Er$^{3+}$ ion is $J = 15/2$. The $(2J+1)$ levels associated with this multiplet are split into eight Kramers doublets due to the crystalline electric fields (CEFs) generated predominantly by the neighboring oxygen ions. To reduce the number of CEF parameters in the Hamiltonian to nine, the quantization axis $z$ can be chosen to coincide with the anticipated Ising axis [010], while $x$ and $y$ are assigned to the other local twofold rotation axes [101] and [10$\bar{1}$]. Assuming no crystallographic distortion, the CEF Hamiltonian is written as: 
\begin{equation}
\label{eq1}
{\hat{\cal H}}_{D_2}^{CEF} = \sum_{i=0,2} B_2^{i}\hat{O}_2^{i} + \sum_{i=0,2,4} B_4^{i}\hat{O}_4^{i} +\sum_{i=0,2,4,6} B_6^{i}\hat{O}_6^{i},
\end{equation}
where the $B_n^i$ are the crystal field parameters to be determined experimentally and $\hat{O}_n^i$ are the Stevens operators\cite{hutchings}. For a given set of crystal field parameters, the Hamiltonian in Eq.~(1) can be diagonalized to find the corresponding CEF energy levels and wavefunctions that can be probed directly with inelastic neutron scattering. The unpolarized double differential cross-section for magnetic neutron scattering can be written as follows\cite{squires}:
\begin{equation}\label{eq2}
\frac{d^{2}\sigma} {d\Omega dE_f} = C\frac{k_f} {k_i}f^{2}(Q)S(Q,\hbar\omega),
\end{equation}
where $\Omega$ is the scattered solid angle, $E_f$ is the final neutron energy, $k_{f/i}$ is the scattered/incident momentum of the neutron, $C$ is a constant, $f(Q)$ is the magnetic form factor, and $S(Q,\hbar\omega)$ is the scattering function.
The scattering function provides the relative scattered transition intensities between different CEF levels and is given by\cite{15_gaudet}:
\begin{equation}\label{eq3}
S(Q,\hbar\omega) = \sum_{i,i'}\frac{(\sum_{\alpha}{|\langle i|J_\alpha|i'\rangle|}^2)e^{-\beta E_i}} {\sum_j e^{-\beta E_j}}L(E_i - E_i' + \hbar\omega),
\end{equation}
where $\alpha = x, y, z$, and $L(E_i - E_i' + \hbar\omega)$ is a Lorentzian function ensuring energy conservation as the neutron induces transitions between CEF levels with energies $E_i$ and $E_i'$.

\begin{figure}
\includegraphics[width=\columnwidth]{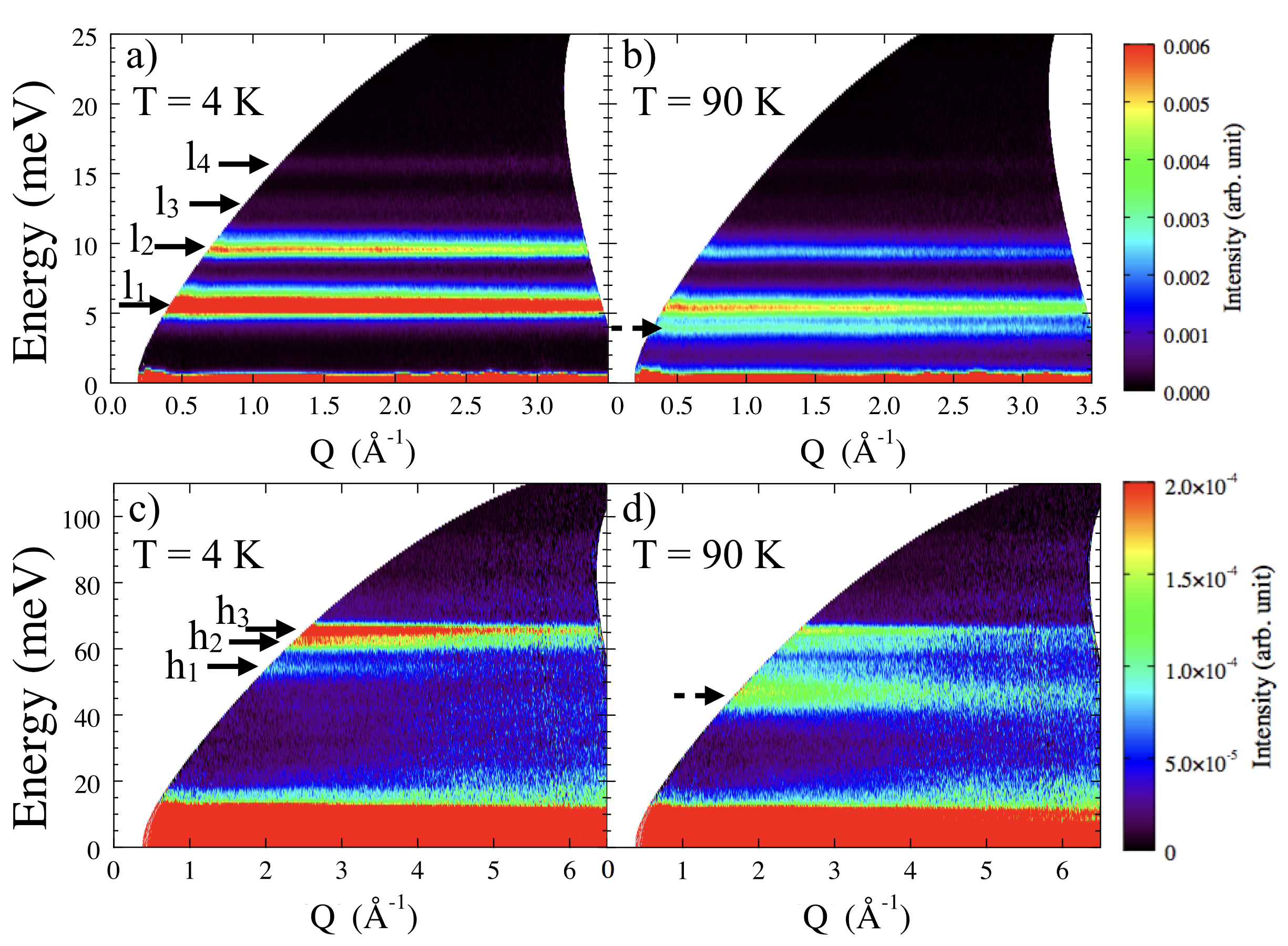}
\caption{Inelastic neutron scattering data for Er$_3$Ga$_5$O$_{12}$ presented as color contour plots, with (a) $E_i$~$=$~30~meV and $T$~$=$~4~K, (b) $E_i$~$=$~30~meV and $T$~$=$~90~K, (c) $E_i$~$=$~120~meV and $T$~$=$~4~K, and (d) $E_i$~$=$~120~meV and $T$~$=$~90~K. Two bands of crystal field excitations are observed in the 4~K data. The lower energy band consists of four modes while the higher energy band consists of three modes; all seven of these excitations are indicated by black arrows in panels (a) and (c). A temperature of 90~K is sufficient to thermally-populate the lowest two crystal field excitations, which generates higher order transitions in these higher-temperature data sets. The higher order transitions are indicated by dashed arrows in panels (b) and (d). }
\label{Fig: INS_spectra}
\end{figure}

\begin{figure}[htb]
\includegraphics[width=\columnwidth]{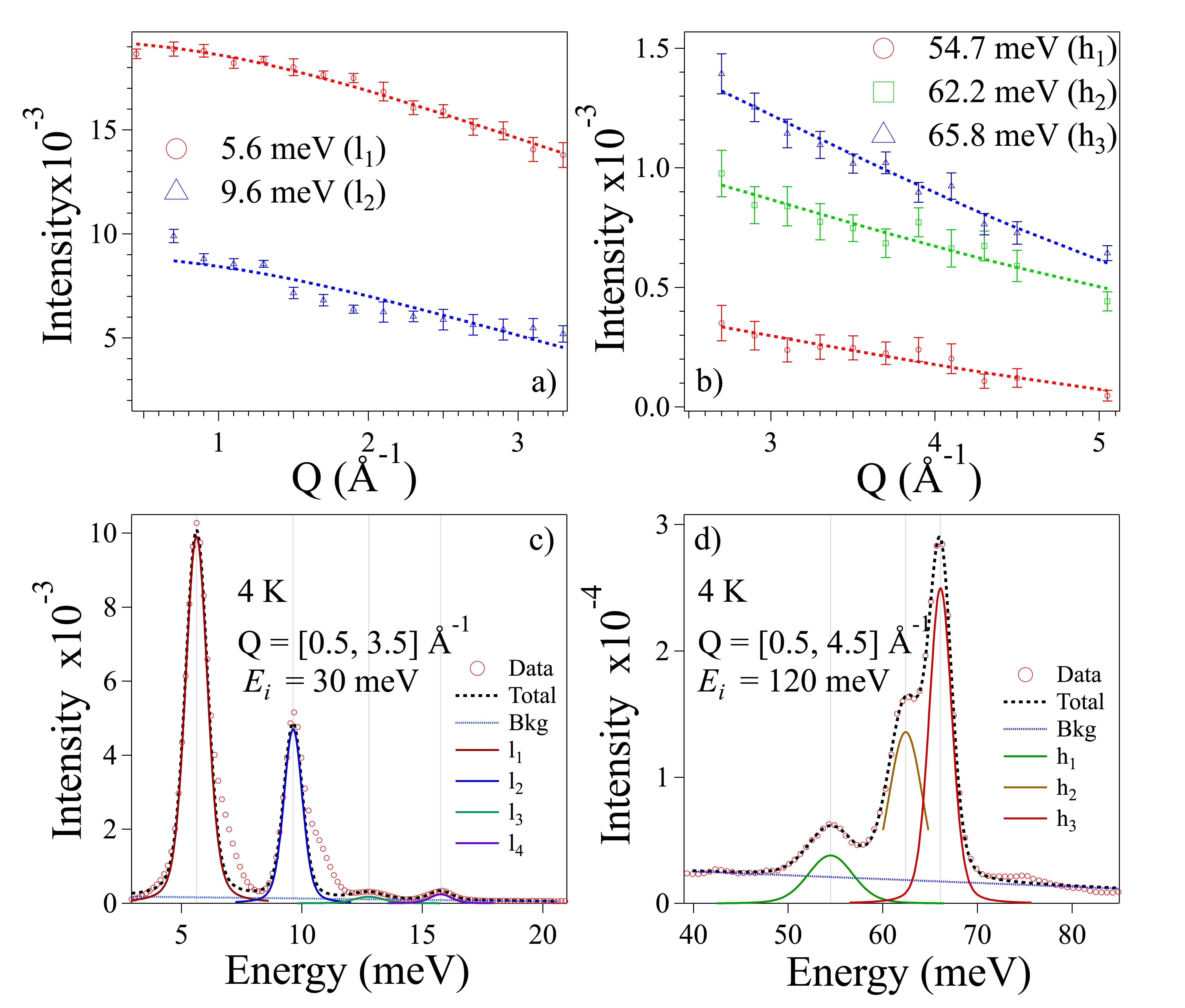}
\caption{(a) The $Q$-dependence of the two most intense CEF excitations from the lower energy band.  (b) The $Q$-dependence of the three CEF excitations from the higher energy band. The intensity of all five of these excitations decreases with increasing $Q$, which is consistent with a magnetic origin. (c) The neutron scattering intensity as a function of energy transfer with $E_i$~$=$~30~meV, revealing the presence of four CEF excitations. (d) The neutron scattering intensity as a function of energy transfer with $E_i$~$=$~120~meV, revealing the presence of three additional CEF excitations. The best fit to these combined datasets, yielding the crystal field parameters presented in Table~II, is indicated by the solid curves in these two panels. }
\label{Fig: formfactor}
\end{figure}

In Fig.~\ref{Fig: INS_spectra}, we present the INS data of Er$_3$Ga$_5$O$_{12}$ collected at $T$~$=$~4 and 90~K with incident energy $E_i$~$=$~30 or 120~meV. As CEF excitations arise from unpaired electrons and therefore have a magnetic origin, they should decrease in intensity with increasing momentum transfer $(Q)$ following the square of the Er$^{3+}$ magnetic form factor. The intensity of CEF excitations should also decrease with increasing temperature. Using these two criteria, we have identified seven possible CEF excitations in the $E_i$~$=$~30 and 120~meV datasets, as indicated by the black arrows in Fig.~\ref{Fig: INS_spectra}(a) and (c). The $Q$-dependence of the five most intense modes, shown in Fig.~\ref{Fig: formfactor}(a) and (b), is consistent with expectations for magnetic scattering. The limited $Q$ range and coarser resolution in the $E_i$~$=$~30 and 120~meV data sets respectively made it difficult to extract a reliable $Q$-dependence of the intensity for the weaker 13 and 16~meV modes, but careful inspection of Fig.~\ref{Fig: INS_spectra}(a) and (b) shows that their intensities decrease with increasing temperature. Therefore, we attribute these modes to the seven excited doublets of the $J$~$=$~15/2 ground state multiplet for Er$^{3+}$. These levels are divided into two distinct bands, with three modes located at much higher energies than the other four. 

\begin{table}[htb]
\caption{The CEF energy levels for Er$_3$Ga$_5$O$_{12}$ from the point charge calculation ($point_{cal}$), the INS experiment ($exp$), and the refined CEF Hamiltonian ($Fitted$). The CEF wavefunctions for the ground state doublet, the $g$ tensor components, and the saturated moment $\mu_{Ising}$ along the local [100] Ising direction obtained from the refined CEF Hamiltonian. }
\label{Table:eigenvalue}
\begin{ruledtabular}
\begin{tabular}{lllllllll}
$point_{cal}$(meV)   & 0 & 9.89& 18.10 & 22.40 & 27.79 & 29.10 & 43.36 &45.13  \\
$exp$ (meV)   & 0 & 5.72 & 9.79 & 13.09 & 16.03 & 53.75 & 61.42 & 65.13   \\
$Fitted$ (meV)       & 0 & 5.78 & 9.78 & 12.92 & 16.07 & 56.08 & 59.43 &64.07 \\
\hline \\[1pt]
\multicolumn{9}{l}{$\begin{aligned}
\phi_0^{\pm} = -0.0068\mid\mp{\frac{13}{2}}\rangle - 0.0490\mid\mp{\frac{9}{2}}\rangle - 0.0335\mid\mp{\frac{5}{2}}\rangle \\
 + 0.0691\mid\mp{\frac{1}{2}}\rangle - 0.1418\mid\pm{\frac{3}{2}}\rangle - 0.6085\mid\pm{\frac{7}{2}}\rangle \\
 - 0.7580\mid\pm{\frac{11}{2}}\rangle + 0.1635\mid\pm{\frac{15}{2}}\rangle
\end{aligned}$} \\ 
\\[1pt] \hline 
\multicolumn{9}{c}{$[g_x, g_y, g_z] = [1.42, 0.03, 11.21]$, $\mu_{Ising} =  \frac{g_z}{2}\mu_B = 5.61 \mu_{B}$}  
\end{tabular}
\end{ruledtabular}
\end{table}

In order to determine the CEF parameters, we began with an initial guess calculated using a point charge model\cite{hutchings} and the known crystal structure. We then diagonalized the Hamiltonian with these parameters to find the corresponding CEF eigenfunctions and eigenvalues, which allowed us to calculate $S(Q,\hbar\omega)$. Finally, we used a least-squares minimization routine with an input of the seven excited CEF energy levels and their corresponding integrated intensities to find a set of nine crystal field parameters that produced the best agreement between the calculated and experimental $S(Q,\hbar\omega)$.

Fig.~\ref{Fig: formfactor}(c) and (d) show constant-$Q$ cuts of the $E_i$~$=$~30 and 120~meV data, indicated by open red circles, with integration ranges of $Q$~$=$~[0.5, 3.5]~\AA$^{-1}$ and [0.5, 4.5]~\AA$^{-1}$ respectively. The best fit curves are superimposed on the data and account for the experimental features quite well, with the exception of the asymmetric peak shapes of the lowest two CEF excitations. We also tried fits that incorporated the full integrated intensity of these two asymmetric CEF modes and found that the refined crystal field parameters were relatively unaffected, which is not surprising since the additional integrated intensity due to the peak asymmetries is quite small. Interestingly, recent work has shown that the instrumental resolution function for a direct geometry spectrometer is asymmetric in energy with a low-energy tail\cite{19_islam}, so the asymmetric broadening observed here appears to come from the sample itself. The lack of additional Bragg peaks in the neutron diffraction data presented below suggests that this broadening cannot be attributed to a magnetic impurity phase. On the other hand, we cannot rule out a scenario where the CEF lineshape asymmetry of these two excitations arises from hybridization with phonon modes\cite{18_gaudet}. Careful single crystal studies may help to elucidate the origin of this broadening.

Table~\ref{Table:eigenvalue} presents the CEF energy levels corresponding to our point charge calculation, the CEF energy levels measured directly by INS, and the CEF energy levels, ground state doublet wavefunctions, g-tensor components ($g_z=2g_J|\langle \phi_0^{\pm}|J_z|\phi_0^{\pm}\rangle|; ~g_{x}, g_{y}=2g_J|\langle \phi_0^{\pm}|J_{x}, -i*J_{y}|\phi_0^{\mp}\rangle|$), and the saturated magnetic moment along the Ising direction obtained from our best fit to the INS data. These g-tensor values determined by our INS data are in broad agreement with those obtained from magnetic susceptibility\cite{67_cooke, 72_redoules}. The crystal field parameters corresponding to both the point charge calculation and our best fit are also shown in Table~\ref{Table:CEF_coefficient}. The ground state doublet wavefunctions and g-tensor components for Er$^{3+}$ are indicative of a strong Ising anisotropy along local [100] directions. 

\begin{table}
\caption{The crystal field parameters for Er$_3$Ga$_5$O$_{12}$ obtained from the point charge calculation and by fitting the INS data at $T$~$=$~4 K.}
\label{Table:CEF_coefficient}
\begin{ruledtabular}
\begin{tabular}{lll}
\multicolumn{1}{c}{$B_n^i (meV)$} & $point_{cal}$ & $Fitted$    \\ \hline
$B_2^0$                         & 0.0779      & 0.1271    \\ 
$B_2^2$                         & -0.2781     & -0.4371    \\
$B_4^0$                           &2.1371e-5  & 6.6574e-4   \\
$B_4^2$                          &-0.0034  & -0.0017                         \\
$B_4^4$                          & -0.0029      & 0.0033                          \\
$B_6^0$                         & -3.0242e-6   & 1.0300e-5                      \\
$B_6^2$                          &-1.4031e-5   & 9.0100e-5                      \\
$B_6^4$                           & 1.9502e-5   & 5.0300e-5                      \\
$B_6^6$                          & -2.1051e-5   & -8.5100e-6                     \\ 
\end{tabular}
\end{ruledtabular}
\end{table}

\subsection{Specific heat and magnetic susceptibility}

\begin{figure}
\includegraphics[width=\columnwidth]{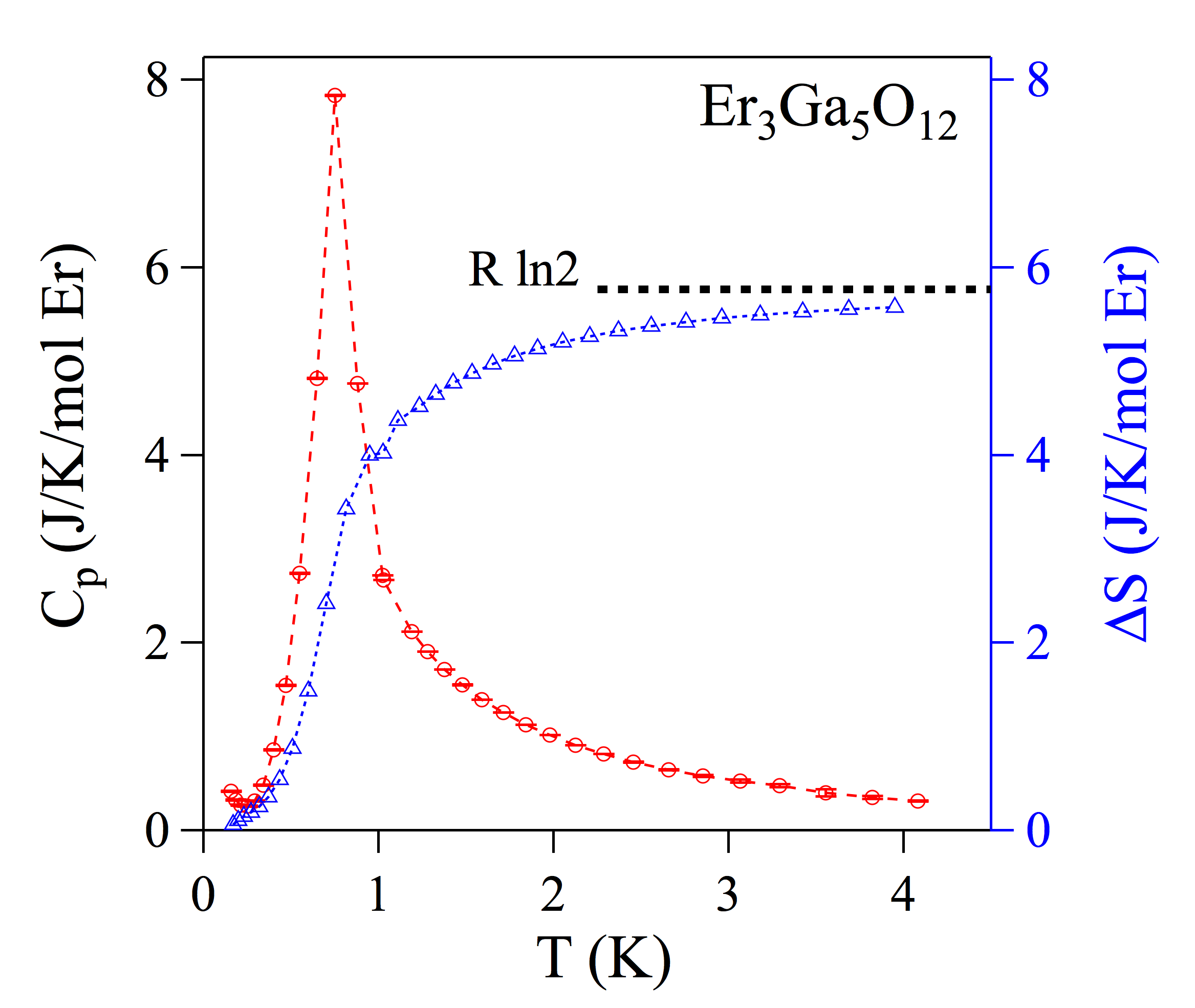}
\caption{Specific heat (red circles, left y-axis) vs temperature for Er$_3$Ga$_5$O$_{12}$ and the entropy recovered (blue triangles, right y-axis) when warming through the ordering transition. Dashed lines are a guide to the eye. The black line shows the expected saturation entropy of $Rln(2)$}
\label{Fig: specific_heat}
\end{figure}

We next turn to the low-temperature collective magnetic properties of Er$_3$Ga$_5$O$_{12}$. We performed a specific heat measurement in zero field to look for signs of magnetic order, low-lying magnetic excitations, and residual entropy. In Fig.~\ref{Fig: specific_heat}, we display our specific heat data for Er$_3$Ga$_5$O$_{12}$ in the low-temperature regime which is in agreement with previous work\cite{67_onn}. The upturn at the lowest temperatures likely arises from a $^{167}$Er nuclear Schottky contribution. With increasing temperature, a second order phase transition is clearly seen around 0.75 K where a $\lambda$ anomaly appears. We also plot the entropy recovery when warming through the ordering transition in Fig~\ref{Fig: specific_heat}, which saturates at $Rln(2)$ by 4~K. This finding is consistent with the well-isolated CEF doublet ground state revealed by INS and indicates that there is essentially no residual entropy remaining in the ordered state. 

\begin{figure}
\includegraphics[width=\columnwidth]{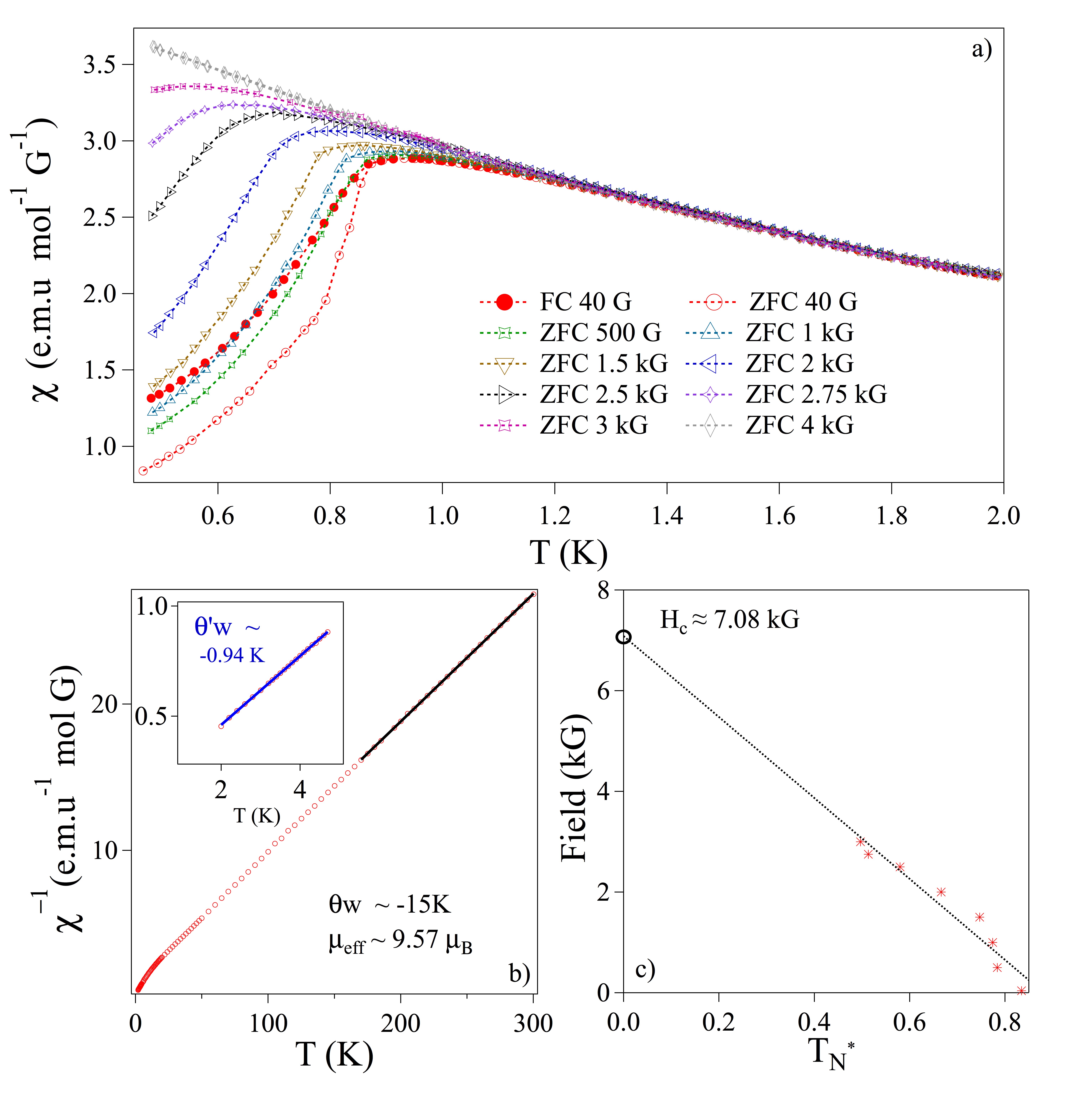}
\caption{(a) Temperature-dependent DC susceptibility below 2 K in selected [110] magnetic fields up to 4 kG. Zero-field-cooled/field-cooled irreversibility was found in the 40~G data only. (b) The inverse susceptibility data (red dots) obtained in a 100~G applied field at temperatures between 2 to 5~K and 150 to 300 K with the Curie-Weiss fits superimposed on the data as solid lines.  (c) External field vs $T_{N^*}$ determined by DC susceptibility measurements. A linear extrapolation of $T_{N^*}$, shown by the dashed line, yields a critical field of 7.08~kG at 0 K. }
\label{Fig: susceptibility}
\end{figure}

As shown in Fig.~\ref{Fig: susceptibility}, the magnetic susceptibility at high temperatures (150-300~K) is well described by a Curie-Weiss Law with a Weiss temperature $\theta_{CW}$~$=$~-15~K and an effective Er moment of 9.57 $\mu_{B}$; the latter corresponds well to the expected value of 9.59 $\mu_{B}$ for an isolated Er$^{3+}$ ion. The Curie-Weiss fit to the magnetic susceptibility in the low temperature regime (2-5~K) results in a Weiss temperature $\theta^{'}_{CW}$~$=$~-0.94~K. Upon cooling, low-field (i.e. 40~G) susceptibility data shows a maximum around 0.8~K where the long-range ordering transition was first identified by zero-field specific heat. The close agreement between $\theta^{'}_{CW}$ and the ordering temperature indicates that Er$_3$Ga$_5$O$_{12}$ is not a strongly-frustrated system. We also found zero-field-cooled/field-cooled irreversibility in the 40~G data only and a strongly field-dependent ordering transition that is suppressed below the temperature range we can measure by 4~kG. A linear extrapolation of the field-dependent ordering temperatures ($T_{N^*}$) gives $H_{c}$(0 K) $\sim$ 7.08~kG, as shown in Fig.~\ref{Fig: susceptibility}(c). 

\subsection{Neutron powder diffraction}

\begin{figure}
\includegraphics[width=\columnwidth]{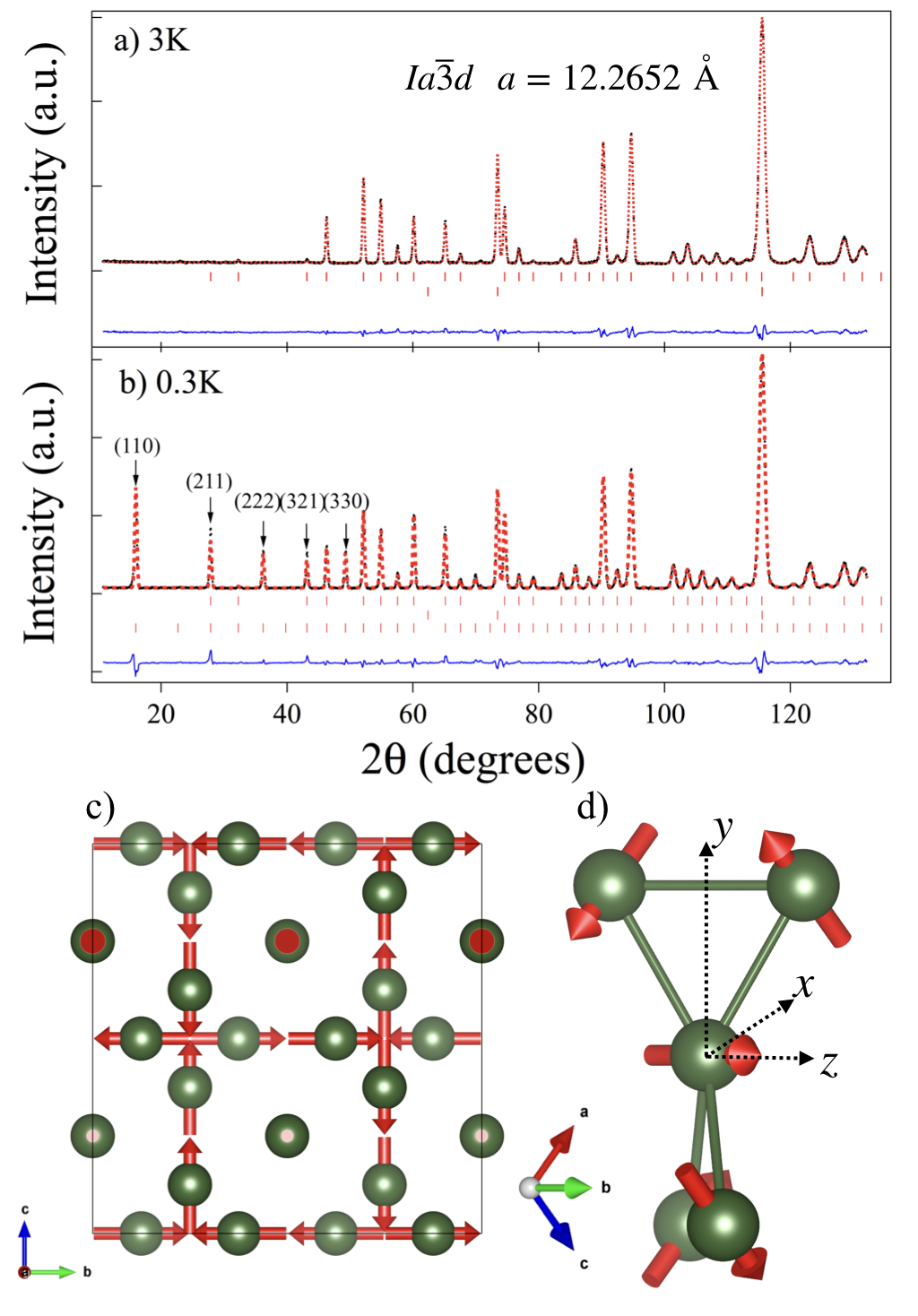}
\caption{Neutron powder diffraction data, indicated by the solid symbols and collected with a neutron wavelength~2.41~\AA, at temperatures of (a) 3~K and (b) 0.3~K. The best structural refinement, including contributions from both the sample and the Al sample can, is superimposed on the 3~K data as a dashed red curve. The magnetic contribution is also included in the 0.3~K refinement. The expected Bragg peak positions for the crystal structure of Er$_3$Ga$_5$O$_{12}$, the crystal structure of the Al can, and the magnetic structure of Er$_3$Ga$_5$O$_{12}$ are indicated by ticks. (c) A schematic of the magnetic structure for Er$_3$Ga$_5$O$_{12}$ as viewed along the a-axis. (d) The spin configuration for the Er$^{3+}$ ions on two corner-sharing triangles.}
\label{Fig: HB2A_final}
\end{figure}

Figure~\ref{Fig: HB2A_final} shows $\lambda$~$=$~2.41~\AA~NPD data for Er$_3$Ga$_5$O$_{12}$ at $T$~$=$~3~K and 0.3~K. The crystal structure of the 3~K data set refines well in the garnet room temperature space group {\it Ia{$\overline3$}d} with lattice constant $a = 12.2652(1)$~\AA, which indicates that cubic symmetry is preserved over a wide temperature range. The 0.3~K NPD data shows evidence for long-range magnetic order, as several new Bragg peaks emerge and some nuclear peaks gain additional intensity. On the other hand, no diffuse magnetic scattering is observed in this data. The five strongest magnetic Bragg peaks are marked by black arrows in Fig.~\ref{Fig: HB2A_final}(b) and can be indexed as (110), (211), (222), (321) and (330); the latter is indicative of a $\vec{k}$~$=$~0 magnetic propagation vector. The representational analysis software SARAh\cite{00_wills} was used to establish candidate magnetic models allowed by symmetry. There are eight possible irreducible representations that describe the spin configurations of this six-sublattice magnet. For clarity, the fractional coordinates for one of the four Er$^{3+}$ ions in the chemical unit cell making up each of the six sublattices are presented in Table~\ref{Table:Fractional_coordinates}. $\Gamma_3$ and $\Gamma_4$ are the simplest magnetic models with only one basis vector each and they are appropriate for Ising garnets with equal moments on all Er$^{3+}$ sites with strong Ising anisotropy along local [100] directions. $\Gamma_5$ and $\Gamma_6$ consist of two basis vectors each and they are also appropriate for Ising garnets with strong Ising anisotropy along local [100] directions, but the moment size on all magnetic sites is not equivalent. The other four models are much more complicated and consist of several basis vectors. 

\begin{table}
\begin{center}
\caption{Fractional coordinates of the Er$^{3+}$ ions corresponding to the six different magnetic sublattices in the {\it Ia{$\overline3$}d} crystallographic unit cell.} 
\label{Table:Fractional_coordinates}
\begin{tabular}{ccc}
\hline \hline
Sublattice & Fractional Coordinates & Moment Direction \\
1 & (0, 0.25, 0.125) & $-\hat{c}$ \\  
2 & (0.125, 0, 0.25) & $-\hat{a}$ \\
3 & (0.25, 0.125, 0) & $-\hat{b}$ \\
4 & (0, 0.25, 0.625) & $\hat{c}$ \\
5 & (0.625, 0, 0.25) & $\hat{a}$ \\ 
6 & (0.25, 0.625, 0) & $\hat{b}$ \\ 
\hline \hline
\end{tabular}
\end{center}
\end{table}

We tried to refine the NPD data using all eight possible models and including both the $j_0$ and $j_2$ spherical Bessel contributions to the Er$^{3+}$ magnetic form factor. We found that the best fit comes from the $\Gamma_3$ magnetic structure, in good agreement with previous work\cite{68_hammann_2}. The final result is superimposed as a dashed red curve on the 0.3~K data presented in Fig.~\ref{Fig: HB2A_final}(b). A schematic of the $\Gamma_3$ magnetic structure viewed along the a-axis is also illustrated in Fig.~\ref{Fig: HB2A_final}(c), while the local arrangement of the Er$^{3+}$ ions on two corner-sharing triangles is shown in Fig.~\ref{Fig: HB2A_final}(d). All the moments in this magnetic structure point along local [100] directions, with the exact configuration specified in Table~\ref{Table:Fractional_coordinates}, and the refined ordered moment is 5.24(4)~$\mu_B$. Both the moment direction and magnitude agree well with our crystal field analysis described above, as we find a slightly larger saturated moment along the local [100] Ising direction of 5.61~$\mu_B$. It is also interesting to note that the $\Gamma_3$ state is the expected magnetic structure for an Ising garnet with moments constrained along the [100] directions and coupled through dipolar interactions only\cite{65_capel}. Furthermore, this is the same spin configuration found for other Ising garnets including Dy$_3$Al$_5$O$_{12}$\cite{65_hastings}, Tb$_3$Al$_5$O$_{12}$\cite{69_hammann}, Ho$_3$Al$_5$O$_{12}$\cite{69_hammann}, and Ho$_3$Ga$_5$O$_{12}$\cite{77_hammann}. 

\subsection{Muon spin relaxation}

In $\mu$SR measurements, spin polarized muons are implanted into a sample one at a time where they thermalize rapidly in the material while maintaining their polarization. These thermalized muons find a minimum electrostatic potential site where they come to rest and their spins precess in the local magnetic field until they decay (with an average lifetime $\tau_{\mu}=2.2~{\mu}s$), emitting a positron preferentially in the direction of the muon spin at the time of decay. Detectors on either side of the sample register the decay of the positron and record the time interval between muon injection and decay.

\begin{figure}
\includegraphics[width=\columnwidth]{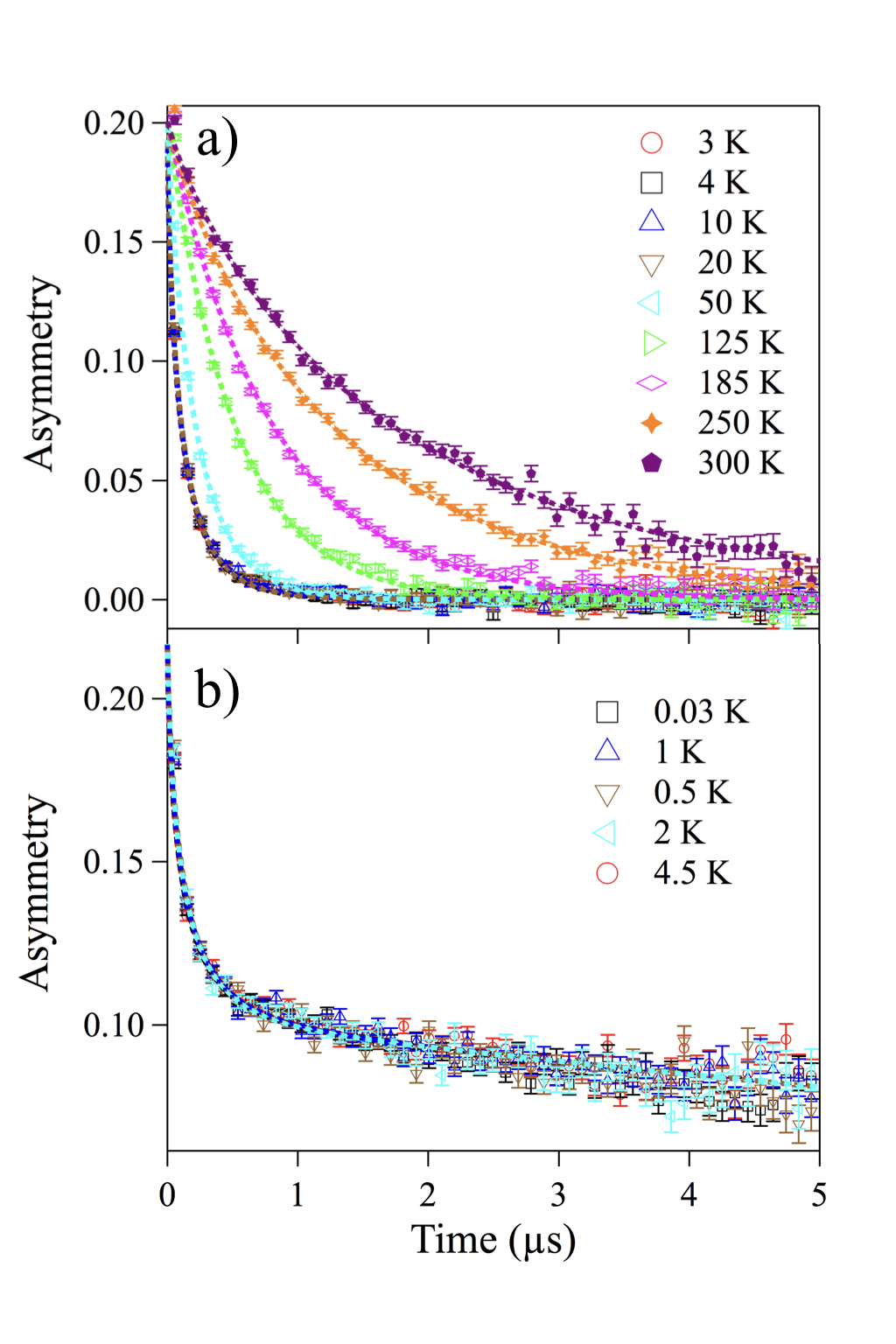}
\caption{Zero field $\mu$SR spectra of Er$_3$Ga$_5$O$_{12}$ measured over a temperature range of (a) 2 to 300 K and (b) 30 mK to 4.5 K. Colored symbols are the experimental data and the dashed lines are the fitting results as described in the text.}
\label{ZFspectra}
\end{figure}

We show ZF$-\mu$SR data for Er$_3$Ga$_5$O$_{12}$ between 30 mK and 300 K in Fig.~\ref{ZFspectra}. This temperature range ensures that the sample passes through the 0.8~K transition temperature inferred from the bulk characterization measurements. We note that the data exhibit no sign of oscillations down to 30 mK, which is unexpected in a magnetically-ordered state. In most cases, no spontaneous muon precession indicates an absence of coherent long-range magnetic order. One alternative scenario for the absence of oscillations is that the initial muon beam polarization and the local field are parallel, but this cannot be the case here due to the non-collinear nature of the magnetic structure as determined by NPD above. Another scenario is that the ordered moment size is too large to resolve in our ZF-$\mu$SR measurements, but we would then expect missing initial asymmetry and we find no evidence for that here. 

\begin{figure}[ht]
\includegraphics[width=\columnwidth]{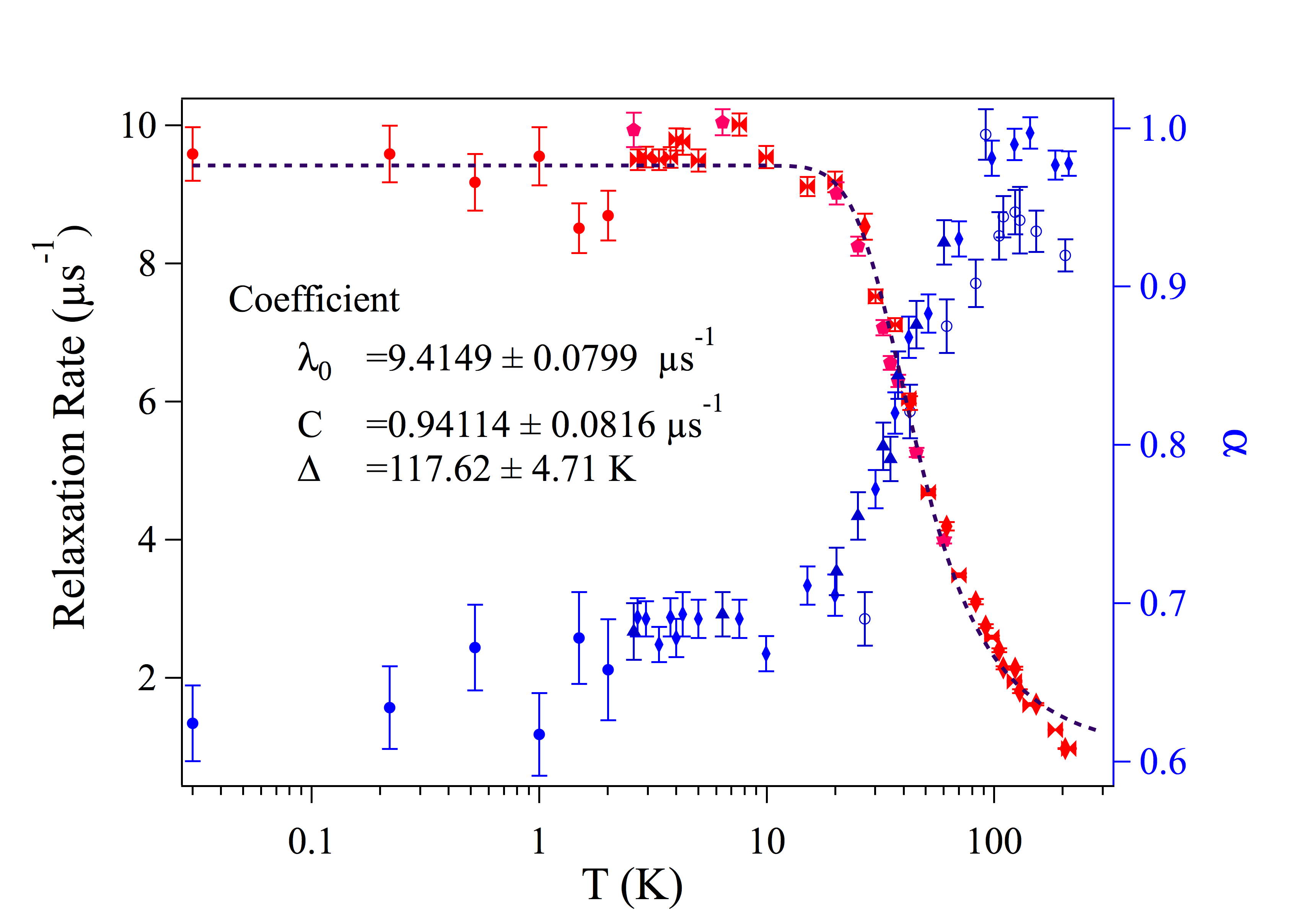}
\caption{Temperature dependence of the zero field $\mu$SR relaxation rate (red symbols) and power $\alpha$ (blue symbols) for Er$_3$Ga$_5$O$_{12}$. The dashed black curve represents the best fit result of the temperature-dependent relaxation rate using the fitting function described in the main text.}
\label{Fig:relaxation_rate}
\end{figure}

We fit the ZF-$\mu$SR spectra at all temperatures to a stretched exponential function of the form: $P(t) = A_{t} {e^{-(\lambda t)^\alpha}}$, with a temperature-independent initial asymmetry $A_{t}$. The temperature-dependence of the relaxation rate $\lambda$ and the power $\alpha$ are shown in Fig.~\ref{Fig:relaxation_rate}. Surprisingly, there is no change in either of these parameters at $T_N$~$=$~0.8~K and they appear to be temperature-independent below $\sim$ 20~K. Similar low-temperature relaxation plateaus have been observed in a variety of frustrated magnets with ordered ground states\cite{05_lago, 06_dalmas, 06_dunsiger, 10_kalvius}, but there is still no consensus on their microscopic origins\cite{11_mcclarty}. In some cases, the presence of both magnetic Bragg peaks in neutron diffraction and relaxation plateaus in ZF-$\mu$SR has been attributed to a dynamical magnetic ground state with a relatively long correlation time\cite{06_dalmas}; we cannot rule out this possibility here.

We also note that ZF-$\mu$SR could be completely insensitive to the ordering transition in this material if the muon occupies a crystallographic site where the local field is zero by symmetry. However, an increase in the relaxation rate below $T_N$ in the isostructural material Yb$_3$Ga$_5$O$_{12}$\cite{03_dalmas} indicates that this possibility is unlikely. To strengthen this conjecture, we calculated the expected field distribution on 1~\AA~spherical shells centered on the O$^{2-}$ ions, corresponding to probable stopping sites for the muons\cite{83_holzschuh}, for the magnetic structure of Er$_3$Ga$_5$O$_{12}$ shown in Fig.~\ref{Fig: HB2A_final}(c). Assuming an ordered moment of 5.24~$\mu_{B}$ as determined by our NPD measurements above, we found that the internal fields range from 0.5 kG to 37.0 kG. We also calculated the local field distribution at these same locations assuming that the Er moments are frozen in a completely random spin configuration, which yields local fields varying from 1.3 kG to 40.3 kG. These calculated fields are all much bigger than the average local field ($\lambda_0/\gamma_{\mu} \sim 110$~G) we infer from our ZF-$\mu$SR spectra, which indicates that the insensitivity of this technique to the ordering transition is not simply due to an accidental cancellation of the local field at the muon stopping sites. 

Above $\sim$ 20 K when the sample is in the paramagnetic state, the relaxation rate decreases with temperature up to 300~K. This behavior is likely due to an Orbach process\cite{61_orbach}, where the Er$^{3+}$ magnetic moment relaxes through a real two-phonon process with an excited crystal field level of energy $\Delta$ as an intermediate state. This process has been shown to play an important role in the $\mu$SR spectra of other rare-earth based magnets\cite{03_dalmas, 08_khasanov, 16_lake, Yaouanc_uSR}. In this case, the temperature-dependence of the relaxation rate can best be modeled by $\lambda^{-1} = \lambda_0^{-1} + {C^{-1}e^{-\beta\Delta}}$ where $\beta = 1/{k_BT}$. The fitted result for Er$_3$Ga$_5$O$_{12}$ is presented in Fig.~\ref{Fig:relaxation_rate} and provides a good description of the relaxation rate over a wide temperature range when $\lambda_{0} = 9.41(8)~ \mu s^{-1}$, $C = 0.94(8)~\mu s^{-1}$, and $\Delta = 10.1(4)$ meV. The value for $\Delta$ is in reasonable agreement with the second excited CEF level at 9.79 meV measured with INS. 

\section{Conclusion}
We have carried out a series of comprehensive measurements investigating the single ion and collective magnetic properties of the garnet Er$_3$Ga$_5$O$_{12}$. Our inelastic neutron scattering measurements reveal a CEF Hamiltonian for Er$^{3+}$ that is consistent with a large Ising anisotropy along local [100] directions. Our bulk characterization measurements, including specific heat and magnetic susceptibility, show evidence for the onset of long-range antiferromagnetic order at 0.8~K. While no evidence for the ordered state is observed in ZF-$\mu$SR, possibly due to a finite correlation time or the signal being dominated by an Orbach process, neutron powder diffraction reveals a six-sublattice, Ising antiferromagnetic spin configuration. This spin configuration is consistent with predictions for [100] Ising moments coupled through dipolar interactions\cite{65_capel} and ensures that Er$_3$Ga$_5$O$_{12}$ is an excellent model system for investigating the complex metamagnetic behavior expected for a multi-axis magnet. 

\section{Acknowledgments}
We acknowledge useful discussions with Connor Buhariwalla, Alannah Hallas and Jonathan Gaudet. We also appreciate the support of TRIUMF personnel during the $\mu$SR measurements. Work at McMaster was supported by the Natural Sciences and Engineering Research of Council of Canada. A portion of this research used resources at the Spallation Neutron Source and High Flux Isotope Reactor, which are DOE Office of Science User Facilities operated by Oak Ridge National Laboratory.

\end{document}